\documentclass[fleqn,twoside,twocolumn,nofootinbib,showkeys]{revtex4} 
\usepackage[nocpr]{ujp} 

\usepackage{dcolumn}
\usepackage{bm}

\begin{document}
\title[$\Lambda$-anomaly in the hadronic chemical freeze-out]
{\boldmath$\Lambda$-ANOMALY\\ IN THE HADRONIC CHEMICAL FREEZE-OUT}
\author{V.V. Sagun}
\affiliation{Bogolyubov Institute for Theoretical Physics, Nat.
Acad. of Sci. of Ukraine}
\address{14-b, Metrolohichna Str., Kyiv, 03680, Ukraine}
\email{v\_sagun@ukr.net}
 \udk{???}  \razd{}

\autorcol{V.V.\hspace*{0.7mm}Sagun}

\setcounter{page}{755}%

\begin{abstract}
A new way to overcome the $\Lambda$ hyperon selective suppression,
which is known as the $\Lambda$-anomaly, has been suggested.\,\,In
particular, the additional radius of a $\Lambda$ hyperon is
introduced into the model of hadron resonance gas with the
multicomponent hard-core repulsion.\,\,The proposed approach allows
one to describe the hadron multiplicity ratios measured at the AGS,
SPS and RHIC energies with the accuracy
$\chi^{2}/dof=$~52/55~$\simeq $~0.95.
\end{abstract}
 \maketitle

\section{Introduction}

\label{Introduction}

A number of extremely complicated experiments devoted to collisions
between heavy ions and aimed at studying the properties of a
strongly interacting matter under extreme conditions were carried
out during last decades.\,\,For instance, the main objective of
experiments carried out on the RHIC and LHC accelerators is the
research of the properties of a strongly interacting matter under
extreme conditions and searches for a new state of the QCD matter, a
quark-gluon plasma.\,\,The theoretical efforts made in the framework
of the lattice QCD and phenomenological approaches are also aimed at
constructing the corresponding phase diagram.\,\,Combining together
hydrodynamic, kinetic and thermal models~-- each of them cannot
describe the collision process completely, but only one of its
stage~-- all stages of the system evolution can be
reproduced.\,\,For instance, the stage of chemical freeze-out (CF)
is described by the hadron resonance gas model (HRGM)
\cite{Thermalmodelreview, KABAndronic:05, KABugaev:Horn2013}.\,\,The
authors of cited works consider the CF as a stage of collisions
between heavy ions, at which all particles do not interact with one
another inelastically, so that the hadron multiplicities can change
only owing to decays.\,\,In addition, the HRGM is based on the
assumption of a thermal equilibrium in the system.\,\,Therefore, the
hadron yields are completely governed by the equilibrium parameters
of CF, namely, the temperature $T$ and the chemical potentials
$\mu_{B}$, $\mu_{S}$ and $\mu_{I_3}$, which correspond to the
conservation of the baryon charge, strangeness and third isospin
projection, respectively.\,\,The relevance of this approach to the
description of particle yields was demonstrated for the collision
energies in the interval from AGS to LHC \cite{Thermalmodelreview,
KABAndronic:05, KABugaev:Horn2013, BugaevEPL:13, KABugaev:Stachel}.

It should be noted that, till now, the HRGM had substantial problems
concerning the description of hadrons containing (anti)strange
quarks.\,\,In turn, this circumstance did not make it possible to
describe the available experimental data with a high accuracy.\,\,It
is especially actual for the dependence of such multiplicity ratios
as $K^{+}/\pi^{+}$ (the Strangeness Horn) and $\Lambda/\pi^{-}$ on
the collision energy.

Traditionally, in order to improve the description of all strange
hadrons, the strangeness suppression factor $\gamma_s$ is used; it
was proposed in work \cite{Rafelski:gamma} as a free parameter in
simulations.\,\,This parameter describes a deviation of the
(anti)strange hadrons from the chemical equilibrium in the system.
However, it turned out that the inclusion of the parameter
$\gamma_{s}$ into the model with the one-component hard-core
repulsion did not give rise to a substantial improvement of the
description of hadron multiplicities \cite{Becattini:gammaHIC,
PBM:gamma}.\,\,Just the application of the HRGM with multicomponent
repulsion \cite{BugaevEPL:13} made it possible to describe the
experimental data with a high accuracy ($\chi^{2}/dof\simeq1.06$)
and demonstrate the importance of the strangeness suppression factor
application.\,\,Moreover, unlike previous results
\cite{Becattini:gammaHIC}, the strangeness enhancement rather than
its suppression was revealed at low energies \cite{BugaevEPL:13,
UJPGamma}.\,\,Although the approach on the basis of the parameter
$\gamma_{s}$ has already been discussed for more than 20 years, its
physical meaning was found only recently in works
\cite{BugaevEPL:13, BugaevIndia, UJPSFO} using the approach of
separated CFs for strange and non-strange hadrons.\,\,As was shown
in work \cite{UJPGamma}, if only one CF is considered for all
hadrons in the system, we will inevitably obtain $\gamma_{s}>1$.
This result really testifies to the absence of a chemical
equilibrium between strange and all other hadrons, especially at low
collision energies, as well as to the necessity of introducing the
separate freeze-outs for them.

Nowadays, the application of $\gamma_{s}$ fit remains the simplest
and the most effective way to describe the available data.\,\,In
addition, it describes the Strangeness Horn more acccurately.
Therefore, in this work, just the approach developed on the basis of
the free parameter $\gamma_{s}$ is applied.

As was shown in work \cite{KABugaev:Horn2013}, a multicomponent
repulsion of the hard-core type is a necessary element of the
HRGM.\,\,The introduction of the corresponding radii of hadrons
takes the repulsion between constituents into account, whereas the
attraction between them is taken into consideration with the help of
many sorts of hadrons.\,\,An adequate way of introducing the
hard-core repulsion is the consideration of hadron gas as a
multicomponent mixture of particles with different radii
\cite{KABugaev:Horn2013, BugaevEPL:13, BugaevtwocompVdW,
MultiComp:12}.\,\,In order to provide the best description for all
data, the baryon, $R_{b}$ and meson, $R_{m}$, radii were fixed in
work \cite{BugaevEPL:13} at values of 0.2 and 0.4~fm, respectively,
whereas the radii of kaons, $R_{K}$ and $\pi $-mesons, $R_{\pi}$,
were fitted independently.\,\,As a result, a high-quality fitting of
experimental data was carried out with the help of the
multicomponent HRGM and the free parameter $\gamma_{s}$, with $\chi
^{2}/dof\simeq1.15$ for 111~independent ratios between hadron
multiplicities measured at 14~values of collision energy in the
center-of-mass system ranging from 2.7 to 200~GeV.\,\,An especially
considerable improvement of the description was obtained for the
energy dependence of the ratio $K^{+}/\pi^{+}$ with the accuracy
$\chi^{2}/dof=3.3/14$ (cf.\,\,with the previous value $\chi
^{2}/dof=7.5/14$ obtained in work~\cite{KABugaev:Horn2013}).

At the same time, the theoretical description of experimental data
for the $\Lambda/\pi^{-}$ and $\bar{\Lambda}/\pi^{-}$ ratios is not
satisfactory.\,\,For instance, one of the best descriptions of the
ratio $\Lambda/\pi^{-}$ was carried out in work \cite{BugaevSFO13}
with the accuracy $\chi^{2}/dof=10/8$.\,\,Difficulties with the
description of the ratios that include the $\Lambda$ and
$\bar{\Lambda}$ hyperons are not new.\,\,As was marked in works
\cite{KABAndronic:05, KABAndronic:09, KABugaev:Horn2013,
BugaevSFO13}, a too slow decrease of the model data description for
the ratio $\Lambda/\pi^{-}$ is a consequence of the
$\bar{\Lambda}$-anomaly, which was detected in works \cite{AGSL3,
KABAndronic:05}. Similar conclusions about the selective suppression
of the yields of the $\bar{p}$, $\bar{\Lambda}$ and $\bar{\Xi}$
multiplicities were also drawn in works \cite{KABugaev:Becattini,
KABugaev:Becattini13, KABugaev:Stachel}.\,\,To solve this problem, I
propose to introduce the hard core radius of the $\Lambda$ hyperon
and, in such a way, to account for the peculiarities of its
interaction in comparison with all other hadrons.\,\,As will be
demonstrated below, this supplement to the HRGM makes it possible to
substantially improve not only the description of the most
problematic ratios between the particle multiplicities, but also the
general description of all other ratios measured in a wide energy
interval from AGS to RHIC.

The structure of the work is as follows.\,\,The next section
contains the basic concepts of the HRGM. Section~\ref{Data} is
devoted to the consideration of experimental data, which are used in
this work.\,\,The obtained fitting results and some speculations
concerning the improvement of the hadron multiplicity description
are presented in Section \ref{Results}.\,\,Section \ref{Conclusions}
with summarizing conclusions ends the paper.\vspace*{-2mm}

\section{Hadron Resonance Gas Model}

\label{Model}

Hadron multiplicities are described by means of the multicomponent
HRGM, which is one of the best thermal models at present.\,\,As was
shown in works
\cite{Thermalmodelreview,KABAndronic:05,Becattini:gammaHIC,KABugaev:Horn2013,BugaevtwocompVdW}%
, the quantum statistics can reasonably be neglected at
corresponding temperatures of the hadron gas and the repulsion
between the constituents can effectively be described with the help
of hard-core radii.\,\,At the same time, the attraction between
hadrons is described, as was done in the statistical bootstrap model
\cite{SBM}, by means of a considerable number of hadronic degrees of
freedom.\,\,For the effective description of hadron multiplicities,
the hard-core radii of pions, kaons, all other mesons, $\Lambda$
hyperons and all other hadrons were fitted.

For the thermodynamic description of the hadronic CF, a Boltzmann
gas consisting of $s$ sorts of hadrons in the volume $V$ and at the
temperature $T$ is considered.\,\,The number of hadrons of the
$i$-th sort will be characterized by the variable $N_{i}$, so that
the total number of particles equals $M=\sum_{i=1}^{s}N_{i}$.\,\,For
any two types $i$ and $j$ of interacting
particles, we introduce the excluded volume $b_{ij}=\frac{2\pi}{3}(R_{i}%
+R_{j})^{3}$, which enters, in turn, the matrix of virial
coefficients $B=(b_{ij})$.\,\,This matrix is symmetric, i.e.
$b_{ij}=b_{ji}$.

The canonical partition function for a mixture of van der Waals
gases with multicomponent repulsion looks like
\cite{BugaevtwocompVdW}
\begin{equation}\label{EqI}
Z_{VdW}(T,V,N_i)=\left[\,\prod_{i=1}^s \frac
{\phi_i^{N_i}}{N_i!}\right]\! \left[V -\frac{N^TBN}{M}
\right]^{\!M}\!\!,
\end{equation}
where $N$ is the column matrix,%
\begin{equation}
N=
\begin{pmatrix}
 N_1 \\
 N_2 \\
... \\
N_s
\end{pmatrix}\!\!,\label{EqII}%
\end{equation}
and $N^{T}$ is the corresponding transposed matrix.\,\,The
one-particle thermal density $\phi_{i}(T,m,g)$ corresponding to the
hadron of the mass $m_{i}$ and
the degeneration factor $g_{i}$ is determined from the equation%
\begin{equation}
\phi_{i}(T)=\frac{g_{i}}{(2\pi)^{3}}\int\exp\left(\!  -\frac{\sqrt{k^{2}%
+m_{i}^{2}}}{T}\!\right)  d^{3}k.\label{EqIII}%
\end{equation}
Each $i$-th sort of hadrons is characterized by the total chemical
potential $\mu_{i}\equiv
Q_{i}^{B}\mu_{B}+Q_{i}^{S}\mu_{S}+Q_{i}^{I_3}\mu_{I_3}$, which is
expressed in terms of the baryon chemical potential $\mu_{B}$, the
chemical potential of the third isospin projection $\mu_{I_3}$, the
strange chemical potential $\mu_{S}$ and the corresponding charges
$Q_{i}^{L}$ ($L=B,S,I_3$).

Since the number of particles does not remain constant at the
collisions of heavy ions, it is necessary to use the grand canonical
ensemble with the
partition function%
\begin{equation}
\mathcal{Z}=\sum_{N_{1}=1}^{\infty}\sum_{N_{2}=1}^{\infty}...\sum_{N_{s}%
=1}^{\infty}\left(\,  \prod_{i=1}^{s}\exp\left[
\frac{\mu_{i}N_{i}}{T}\right]
\!\right)   Z_{VdW}.\label{EqIV}%
\end{equation}

In the thermodynamic limit within the method of maximum term
\cite{LeeYang52}, the partition function (\ref{EqIV}) can be
substituted by the term that gives the largest contribution to
$\mathcal{Z}$. \,\,Let it be the matrix $N^{\ast}$. Then the
pressure in the system will be determined by the
relation%
\[
 p/T=\lim_{V \to \infty}\frac{\mathcal{Z}}{V}=  \]\vspace*{-7mm}
\begin{equation}\label{EqV}
= \lim_{V \to \infty}\frac{1}{V}\ln\left[ \,\prod_{i=1}^s \frac
{A_i^{N_i^*}}{N_i^*!}  \left (\! V-\frac{(N^*)^TBN^*}{ M^*}
\!\right)^{\!\! M^*}\right]\!,
\end{equation}
where $A_{i}=\phi_{i}\exp\left[  \frac{\mu_{i}}{T}\right]  $. In
order to find $N^{\ast}$, let us apply the condition of the function
maximum \mbox{($i=1, ...,
s$),}%
\begin{equation}\label{EqVI}
{\partial \over \partial N_i^*} \left[ \ln\left[ \,\prod_{i=1}^s
\frac {A_i^{N_i^*}}{N_i^*!}  \left (\! V-\frac{(N^*)^TBN^*}{ M^*}
\!\right)^{\!\! M^*}\right]  \right]=0.
\end{equation}
Differentiating this formula and making the substitution
$\xi_{i}=\frac {N_{i}^{\ast}}{V-\frac{N^{\ast T}BN^{\ast}}{M}}$, we
obtain
\begin{equation}\label{EqVII}
\xi_i=A_i \exp\left(\!-\sum_{j=1}^s 2\xi_j
b_{ij}+\frac{\xi^TB\xi}{\sum_{j=1}^s\xi_j}\!\right) \!\!.
\end{equation}
The variable $\xi$ is a column vector of corresponding coefficients
$\xi_{i}$, similarly to Eq.~(\ref{EqII}).

The quantity $T\xi_{i}$ is nothing else but the partial pressure of
hadrons of the $i$-th sort.\,\,Hence, using Eq.\,\,(\ref{EqVII}),
the hadronic density $n_{i}=\frac{N_{i}^{\ast}}{V}$ and the system
pressure $p$ can be expressed as
follows:%
\begin{equation}\label{EqVIII}
p=  T\sum_{i=1}^s \xi_i ,
\end{equation}\vspace*{-5mm}
\begin{equation}
 n^L_i = \frac{ Q_i^L{\xi_i}}{\textstyle
1+\frac{\xi^T {\cal B}\xi}{\sum\limits_{j=1}^s \xi_j}}. \label{EqIX}
\end{equation}
Equations (\ref{EqVII}) and (\ref{EqVIII}) allow one to find the pressure in
the system and Eq.~(\ref{EqIX}) gives the thermal multiplicity of charges $Q_{L}$,
provided that the values of $T$ and $\mu_{i}$ are known.

The key fitting parameters of the model are the temperature $T$, the
baryon chemical potential $\mu_{B}$ and the chemical potential of
the third isospin projection $\mu_{I_3}$, whereas the strange
chemical potential $\mu_{S}$ is determined from the condition that
the total strangeness in the system equals zero.\,\,The dependences
of model parameters on the collision energy are shown in Fig.~1 for
$T$, $\mu_{B}$ and $\mu_{I_3}$.\,\,A more detailed description of
the model can be found in works \cite{KABugaev:Horn2013,
MultiComp:12}.

In this work, we also consider the possible deviation of strange
particles from the equilibrium.\,\,The consideration is carried out
in the framework of the conventional procedure, namely, the
introduction of the strangeness suppression factor
$\gamma_{s}$.\,\,The corresponding mathematical implementation
consists in the following substitution of the one-particle thermal
density
\begin{equation} \label{eq:gamma_s}
\phi_i(T) \to \phi_i(T) \gamma_s^{s_i} ,
\end{equation}
where $s_{i}$ equals the total number of strange valence quarks and
antiquarks.\,\,This is the standard procedure \cite{Rafelski:gamma}
that makes it possible to consider a probable deviation of the
strange charge from the total chemical equilibrium.

\begin{figure}%
\vskip1mm
\includegraphics[width=6.3cm, height=6.0cm]{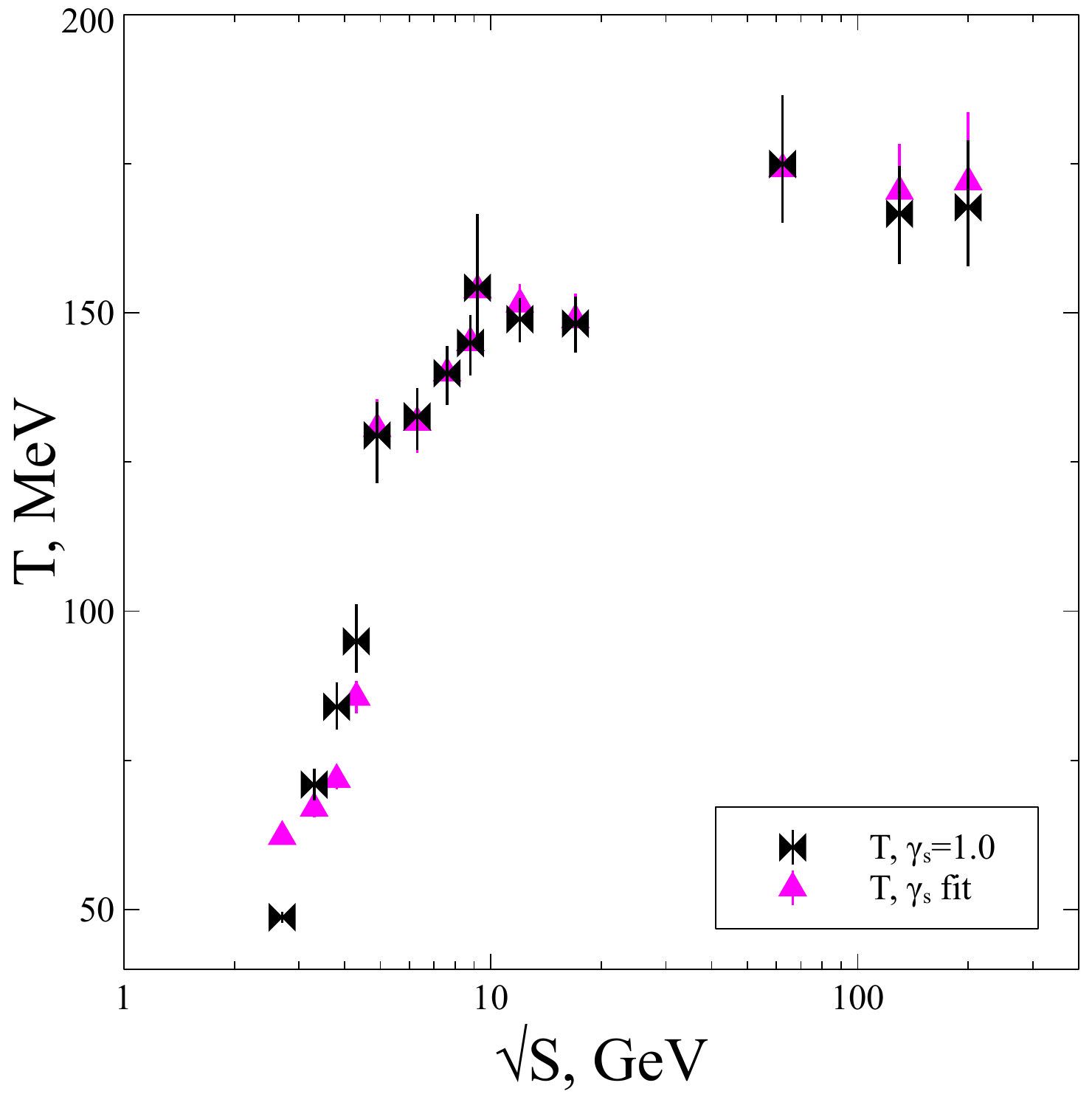}\\[1mm]
\includegraphics[width=6.5cm, height=6.0cm]{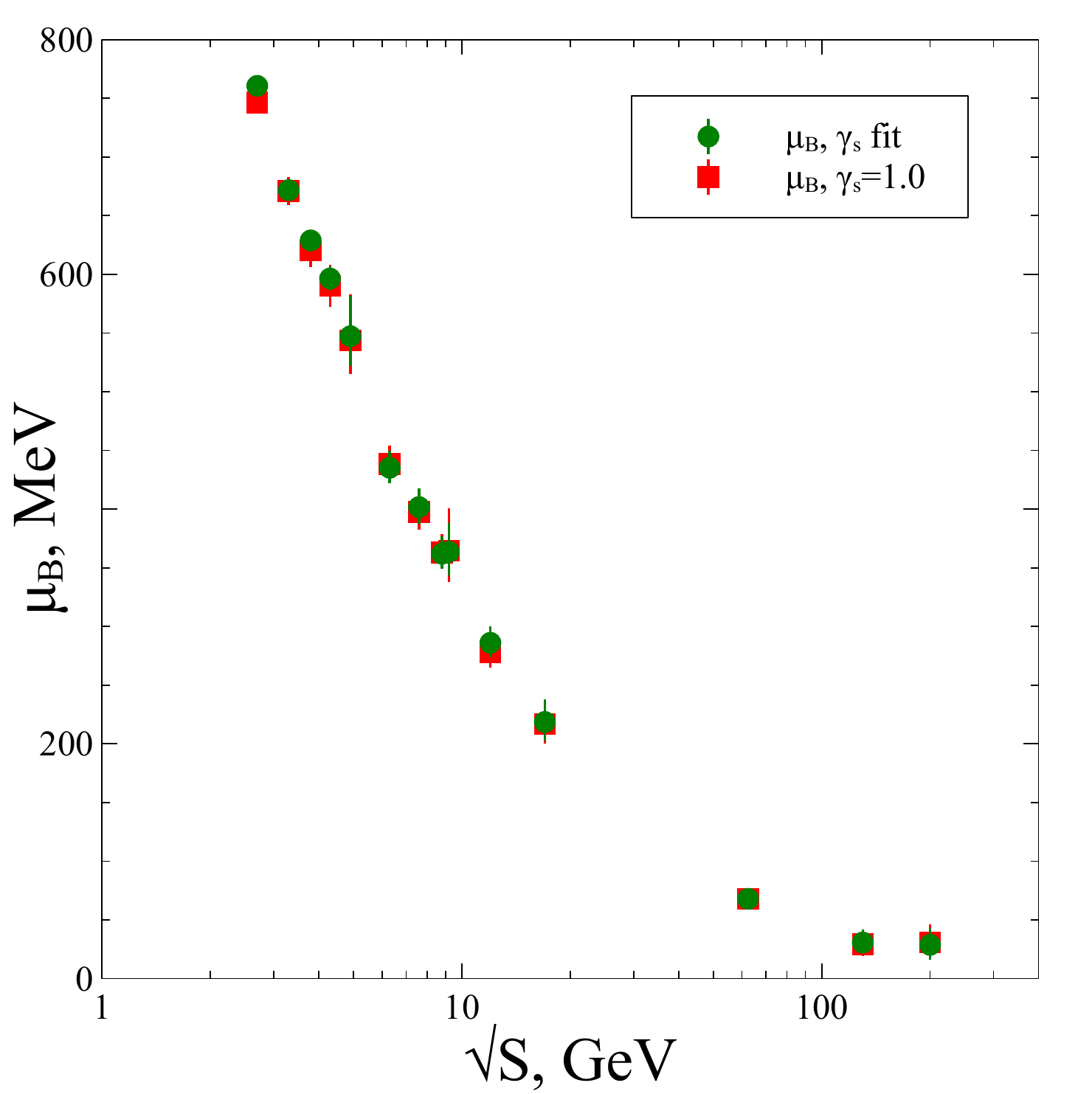}\\[1mm]
\includegraphics[width=6.3cm, height=6.0cm]{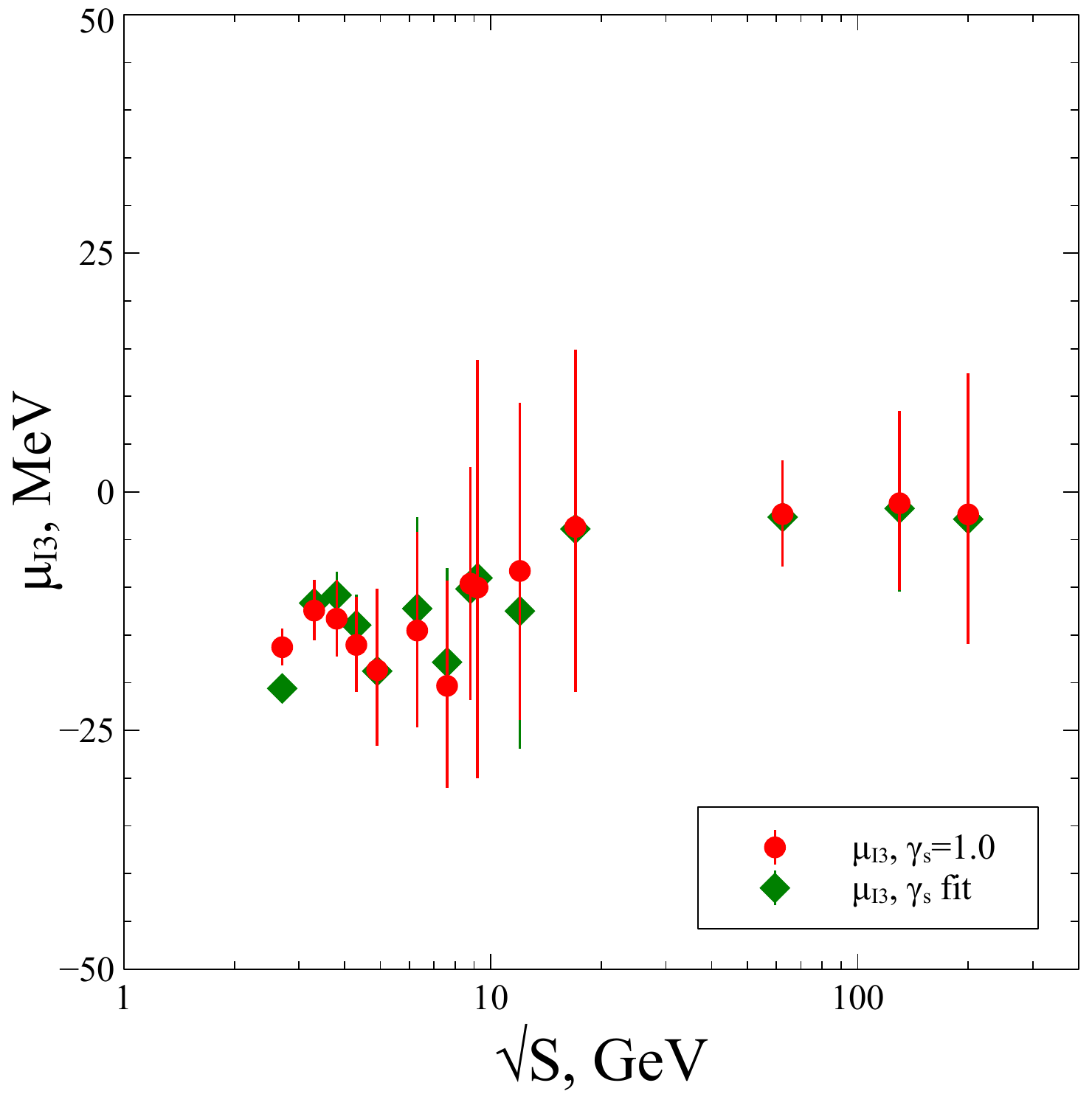}
\vskip-3mm\caption{Behavior of the chemical freeze-out parameters in
the models with the constant ($\gamma_{s}=1$) and fitted parameter
$\gamma_{s}$: the dependences of the chemical freeze-out temperature
(upper panel), the baryon chemical potential (middle panel) and the
chemical potential of the third isospin projection (lower panel) on
the collision energy  }\label{fig_SagunI}
\end{figure}

Making allowance for the width of a hadronic state, $\Gamma_{i}$, is
one of the important elements of this model.\,\,As was demonstrated
in works \cite{KABugaev:Horn2013, Bugaev:1312a}, the thermodynamic
properties of a hadronic system are extremely sensitive to this
width.\,\,Therefore, the finite widths of resonances were introduced
by means of the standard modification of the one-particle thermal
density $\phi_{i}(T)$ \cite{KABAndronic:05}; namely, by averaging
all the terms that include the mass over the Breit--Wigner
distribution function and using the threshold $M_{i}$ for each
resonance.\,\,As a result, the modified one-particle thermal density
for the $i$-th sort of
hadrons takes the form%
\[
\int \exp\left(\! -\frac{\sqrt{k^2+m_i^2}}{T} \!\right)d^3k
\rightarrow  \]\vspace*{-5mm}
\begin{equation}
\label{EqX} \rightarrow \frac{\int^{\infty}_{M_{i}}
\frac{dx}{(x-m_{i})^{2}+\Gamma^{2}_{i}/4} \int
\exp\left(\!-\frac{\sqrt{k^2+x^2}}{T} \!\right)d^3k
}{\int^{\infty}_{M_{i}} \frac{dx}{(x-m_{i})^{2}+\Gamma^{2}_{i}/4}} ,
\end{equation}
where $m_{i}$ is the average hadron mass.

Experimentally detected hadron multiplicities are sums of a thermal
component and a component resulting from hadron decays.\,\,For
instance, many pions appear owing to heavy hadron decays.\,\,The
effect of resonance decays $Y\rightarrow X$ to final hadron
multiplicities is taken into account as
follows:%
\begin{equation}
\label{EqXI} n^{\rm fin}(X) = \sum_Y BR(Y \to X) n^{th}(Y),
\end{equation}
where $BR(Y\rightarrow X)$ is the probability that hadron~$Y$ decays
into hadron~$X$.\,\,In addition, it is supposed for convenience that
$BR(X\rightarrow X)=1$.\,\,All the parameters used in the fitting of
data (the masses $m_{i}$, the widths $\Gamma_{i}$, the degeneration
factors $g_{i}$ and the probabilities of decays for all strong
decay channels) were taken from the particle tables of the
thermodynamic code \textsc{THERMUS} \cite{THERMUS}.

\section{Experimental Data}

\label{Data}

The model described above was used to fit the experimental data on
the collisions between heavy ions, namely, the ratios of particle
yields measured at the midrapidity.\,\,Unlike the particle yield
fit, the fit of particle yield ratio allows us to exclude the
effective volume of the system and, hence, to reduced the number of
model parameters \cite{BugaevEPL:13}.

In particular, high-quality data are available for the AGS collision
energies within the interval $\sqrt{s_{NN}}=$
$2.7$ $\div$ $4.9~\mathrm{GeV}$ (the kinetic beam energy equals from 2
to 10.7~AGeV).\,\,For energies of 2, 4, 6 and 8~AGeV, there are
data on the pion \cite{AGSpi1, AGSpi2}, proton \cite{AGSp1, AGSp2},
and kaon (except for an energy of 2~AGeV) \cite{AGSpi2} yields, as
well as on the $\Lambda$ hyperon yield integrated over the angle of
$4\pi$ \cite{AGSL}.\,\,For a beam energy of 6~AGeV, the
multiplicities of $\Xi^{-}$ hyperon integrated over $4\pi$ were also
measured \cite{AGSKas}.\,\,According to work \cite{KABAndronic:05},
the yields of $\Lambda$ and $\Xi^{-}$ hyperons experimentally
measured at the midrapidity do not meet the requirements; instead,
corrected data from work \cite{KABAndronic:05} are
considered.\,\,For the energy $\sqrt{s_{NN}}=4.9~\mathrm{GeV}$ in
the center-of-mass system, the yields of $\phi$ meson \cite{AGSphi},
and $\Lambda$ and $\bar{\Lambda}$ hyperons \cite{AGSL2, AGSL3} are
also available.\,\,Following work \cite{KABugaev:Horn2013}, I used
the results for particle multiplicities at the midrapidity
experimentally measured by the NA49 Collaboration
\cite{KABNA49:17a,KABNA49:17b,KABNA49:17Ha,KABNA49:17Hb,KABNA49:17Hc,KABNA49:17phi}%
.\,\,Since the results obtained by different collaborations on the
RHIC accelerator for the collisions between high-energy heavy ions
coincide with one another, the STAR Collaboration data for
$\sqrt{s_{NN}}=9.2$~GeV \cite{KABstar:9.2}, 62.4~GeV
\cite{KABstar:62a}, 130~GeV
\cite{KABstar:130a,KABstar:130b,KABstar:130c,KABstar:200a} and
200~GeV \cite{KABstar:200a,KABstar:200b,KABstar:200c} were analyzed.

The minimum of the relative deviation of the fit from experimental data,
$\chi^{2}=\sum_{i}\frac{(r_{i}^{\mathrm{theor}}-r_{i}^{\mathrm{exp}})^{2}%
}{\sigma_{i}^{2}}$, where $r_{i}^{\mathrm{exp}}$ is the experimental
yield of the $i$-th particle, $r_{i}^{\mathrm{theor}}$ the
corresponding theoretically predicted value and $\sigma_{i}$ the
total error of the experimental value, was the criterion of the best
description of experimental data.\,\,In order to determine the
corresponding values of hard-core radii that satisfy the criterion
of root-mean-square deviation minimum, the global fitting of
experimental data was made.\,\,The obtained hard-core radii of
hadrons were fixed and a new fitting procedure was carried out once
more.\,\,Hence, while calculating the number of the degrees of
freedom in the model, the particle radii were not taken into account
(this is a usual practice \cite{Thermalmodelreview, KABAndronic:05,
KABugaev:Horn2013, BugaevEPL:13, KABugaev:Stachel}), because they
were previously determined and fixed.

\section{Fitting Results}

\label{Results}

The best description of experimental data for 14~collision energies
$\sqrt{s_{NN}}=2.7$, 3.3, 3.8, 4.3, 4.9, 6.3, 7.6, 8.8, 9.2, 12, 17,
62.4, 130 and 200~GeV, which correspond to the $\chi^{2}$ minimum,
was obtained at \mbox{$R_{b}=0.355$}~fm, $R_{m}=0.4$~fm,
$R_{\pi}=0.1$~fm, $R_{K}=$ $=0.38$~fm and
$R_{\Lambda}=0.11$~fm.\,\,In addition, a weak dependence of model
parameters on the pion radius was revealed, in contrast to the
variation of the $\Lambda$ hyperon radius.

The fitting of experimental data in the case $\gamma_{s}=1$
testifies to an insignificant improvement of the description,
$\chi^{2}/dof=75.49/69\simeq 1.09$, in comparison with a similar
fitting for four hard-core radii, $\chi^{2}/dof=80.5/69\simeq1.16$
\cite{BugaevEPL:13}.\,\,Let us consider all the most significant
improvements in the description of hadron multiplicities in more
details.\,\,At the collision energies $\sqrt{s_{NN}}=2.7$, 3.3, 3.8,
and 4.3~GeV, the data description made in work \cite{KABAndronic:05}
was already perfect and a further improvement did not take place,
because the number of experimentally measured ratios was equal only
to 4, 5, 5 and 5, respectively; and only kaons and $\Lambda$
hyperons are composed of strange quarks.\,\,For the AGS energy
$\sqrt{s_{NN}}=4.9~\mathrm{GeV}$, the introduction of the additional
radius
$R_{\Lambda}$ considerably improved the description of the ratios $K^{-}%
/K^{+}$ and $p/\pi^{-}$, but worsened the description of the ratios $K^{+}%
/\pi^{+}$ and $\Lambda/\pi^{-}$ (see details in
Fig.~2).\,\,Substantial improvements in the description of $\Lambda$
and $\bar{\Lambda}$ hyperons are observed at the energies
$\sqrt{s_{NN}}= 6.3$, 8.8, 12, 17, 130 and 200~GeV.\,\,One can see
from Fig.~2 that the description quality became considerably higher
for the ratios including $\Lambda$ and $\bar{\Lambda}$ hyperons
($\Lambda/\pi^{-}$, $\bar{\Lambda}/\pi^{-}$ and
$\bar{\Lambda}/\Lambda$).\,\,However, the description of the ratios
that include the kaon yields became worse (not regularly).\,\,For
instance, the quality of the description of the $K^{+}/\pi^{+}$,
$K^{-}/K^{+}$ and $\varphi/K^{+}$ ratios became worse
insignificantly at $\sqrt{s_{NN}}=12~\mathrm{GeV}$.

\begin{figure}%
\vskip1mm
\includegraphics[width=6.0cm, height=5.6cm]{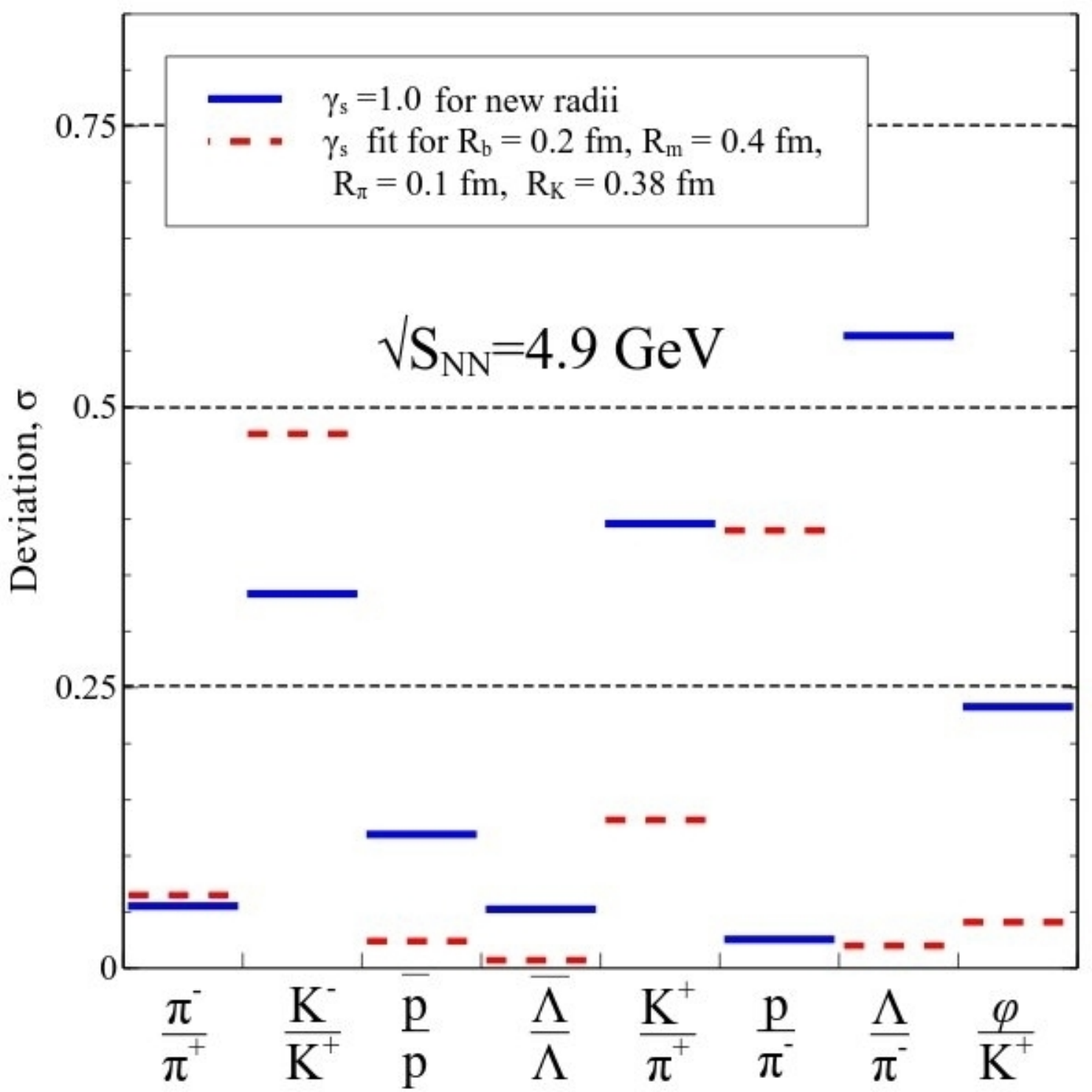}\\[1mm]
\includegraphics[width=6.0cm, height=5.6cm]{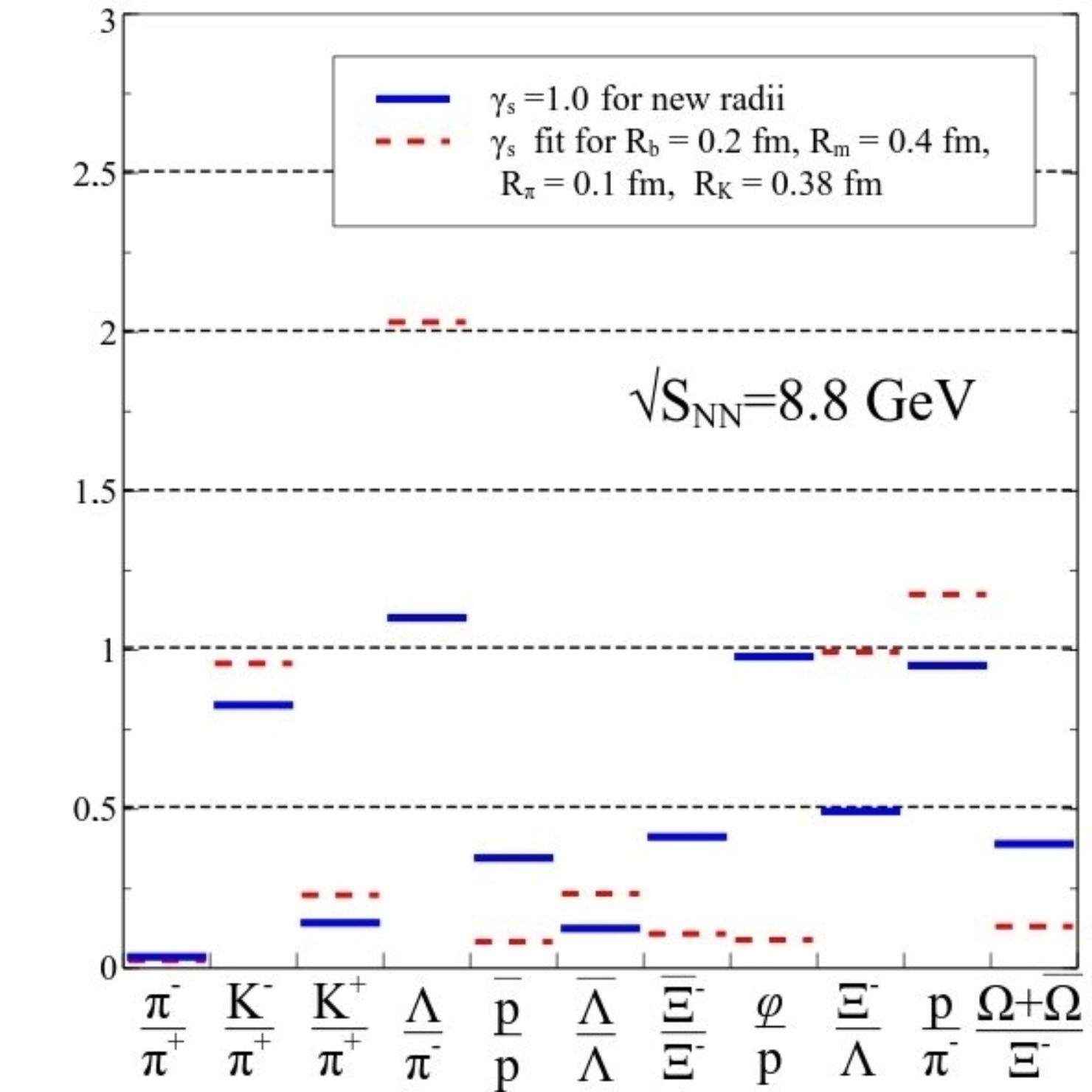}\\[1mm]
\includegraphics[width=6.0cm, height=5.6cm]{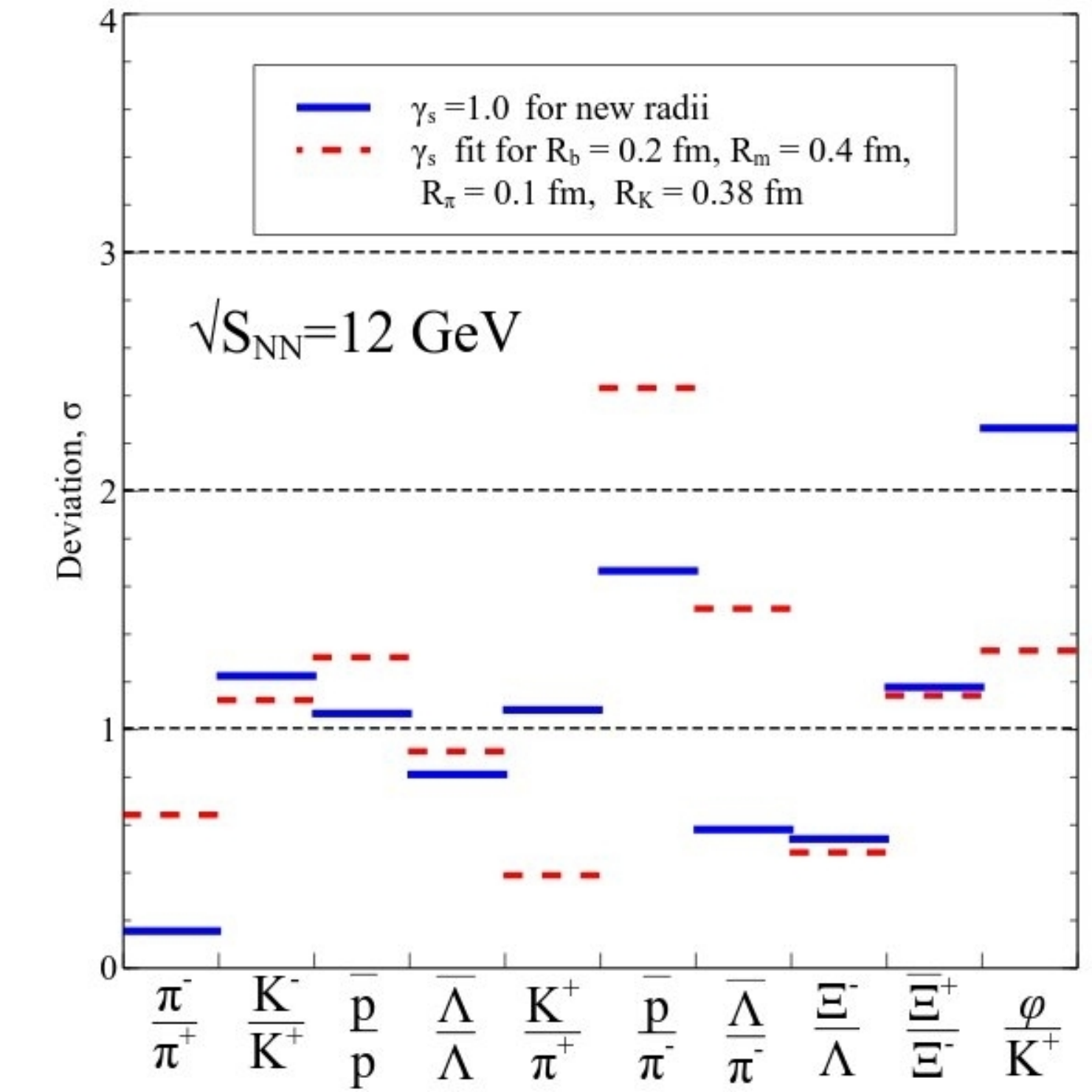}
\vskip-3mm\caption{Relative deviations of the theoretical
description of hadron multiplicities from the corresponding
experimental values reckoned in the experimental error $\sigma$
units for $\sqrt{s_{NN}}=4.9$, 8.8 and 12~GeV (panels from upper to
bottom). The particle ratios are reckoned along the abscissa axis,
and the absolute values of the relative deviation
$\frac{|r^{\mathrm{theor}}-r^{\mathrm{exp}}|}{\sigma^{\mathrm{exp}}}$
along the ordinate one.\,\,The solid lines correspond to
$\gamma_{s}=1$ and the dashed ones to the fit results with the free
$\gamma_{s}$ parameter obtained in work \cite{KABugaev:Horn2013}}
\label{fig_SagunII}
\end{figure}

Regarding $\gamma_{s}$ as a free fitting parameter, the yield
rations can be described better for all particles with the accuracy
$\chi^{2}/dof=52/55\simeq 0.95$.\,\,As is shown in Fig.~3, the
parameter $\gamma_{s}$ exceeds 1 at low energies, which confirms the
strangeness enhancement revealed in work \cite{BugaevEPL:13}.
According to Fig.~3, the dependence of the parameter $\gamma_{s}$ on
$\sqrt{s_{NN}}$ has a local minimum in the energy interval
4.3--4.9~GeV, which may testify to a qualitative variation in the
system properties \cite{Bugaev:1312a}.\,\,The search for
irregularities of this type is important, because they can signals
of the deconfinement phase transition.\,\,For the statement on the
irregularities in this energy interval to be more exact, it is
necessary to carry out a detailed experimental scan in the range of
collision energies from 4 to 5~GeV in the center-of-mass system.

\begin{figure}%
\vskip1mm
\includegraphics[width=6.3cm]{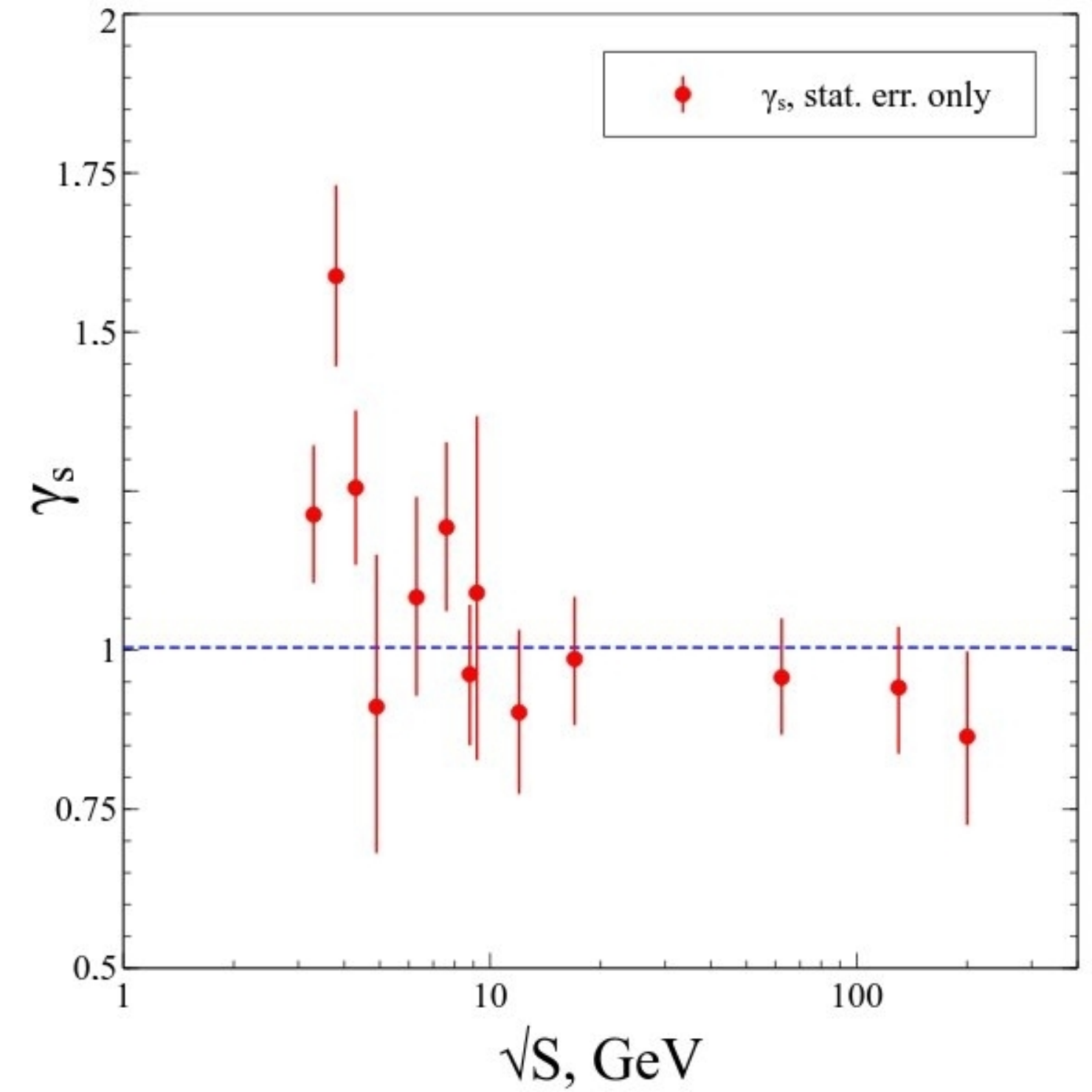}
\vskip-3mm\caption{Dependence $\gamma_{s}(\sqrt{s_{NN}})$
}\label{fig_SagunIII}
\end{figure}

\begin{figure}%
\vskip1mm
\includegraphics[width=6.5cm, height=6.3cm]{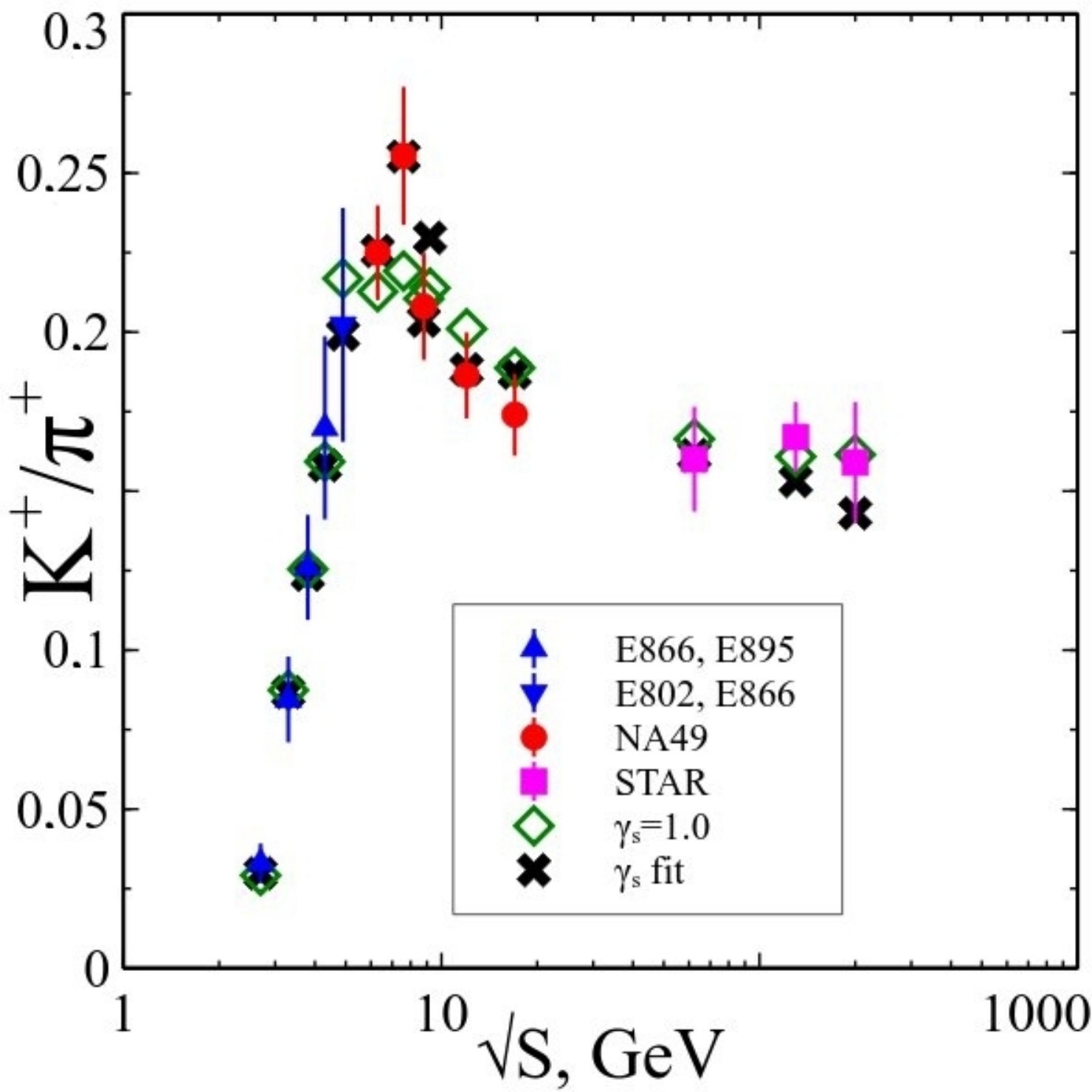}\\[2mm]
\includegraphics[width=6.5cm, height=6.3cm]{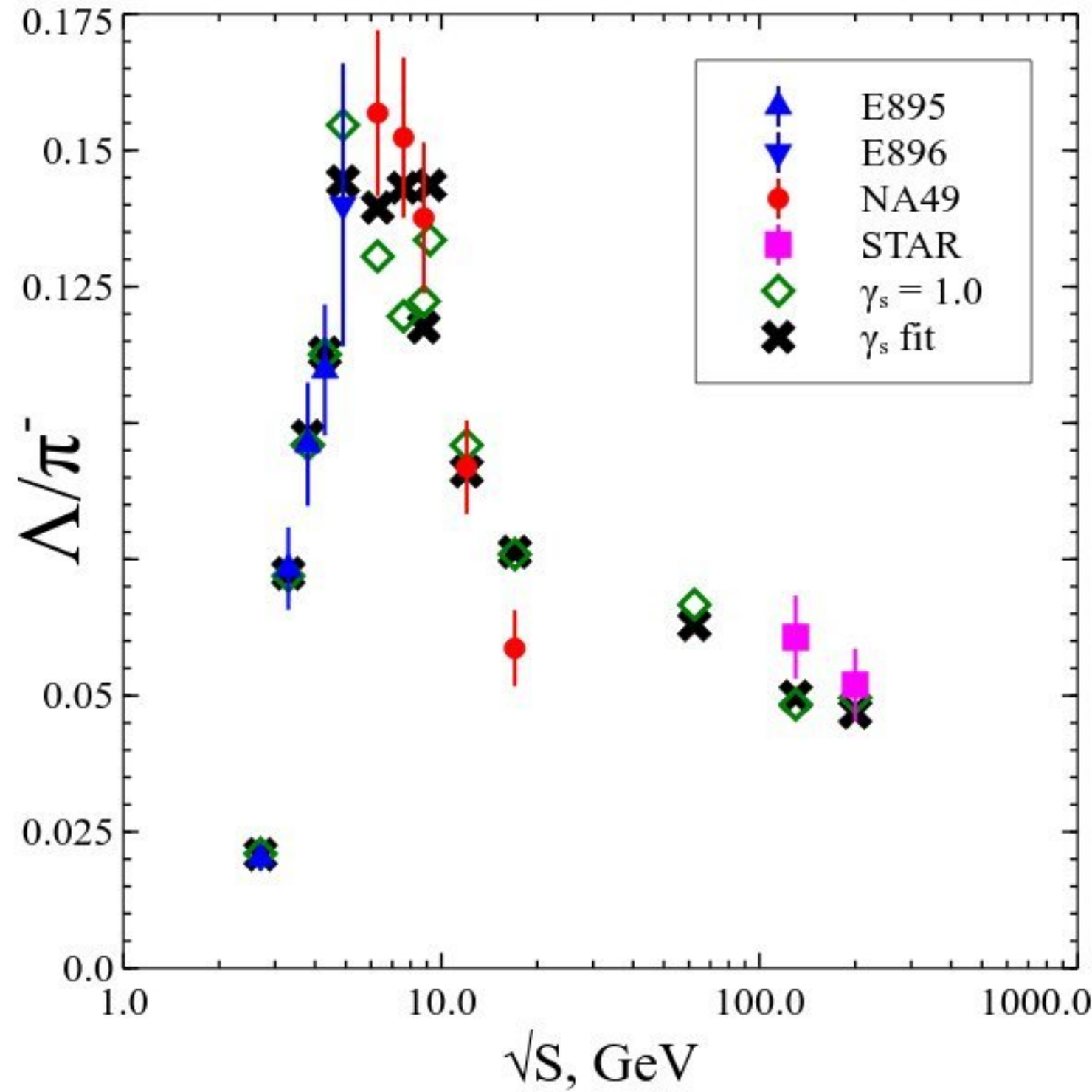}\\[2mm]
\includegraphics[width=6.5cm, height=6.3cm]{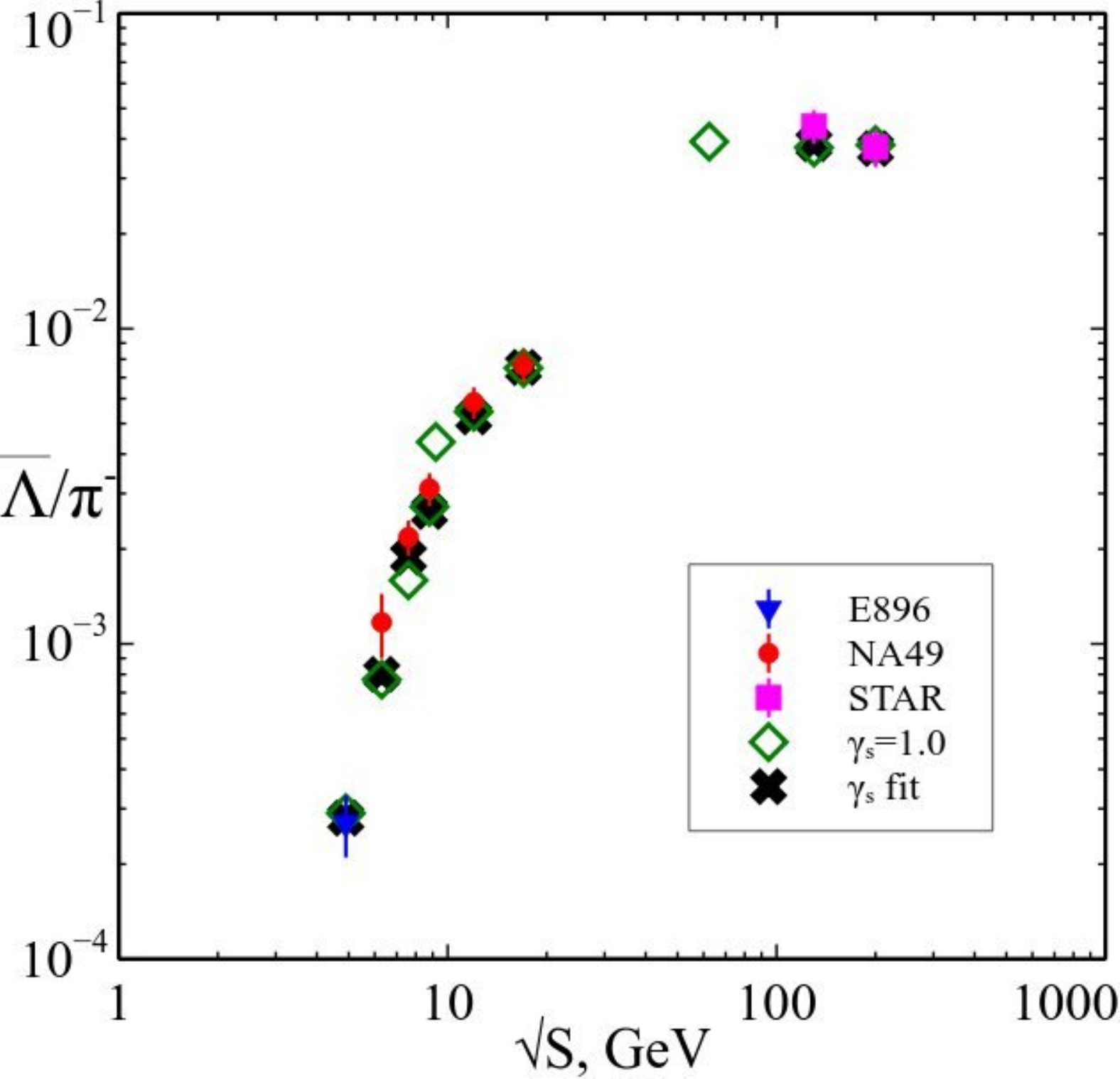}
\vskip-3mm\caption{Dependences of the $K^{+}/\pi^{+}$ (upper panel),
$\Lambda/\pi^{-}$ (middle panel) and $\bar{\Lambda}/\pi^{-}$
(bottom panel) ratios on $\sqrt{s_{NN}}$ for the constant
($\gamma_{s}=1$) and fitted parameter $\gamma_{s}$
}\label{fig_SagunIV}
\end{figure}

\begin{figure}%
\vskip1mm
\includegraphics[width=6.5cm, height=6.15cm]{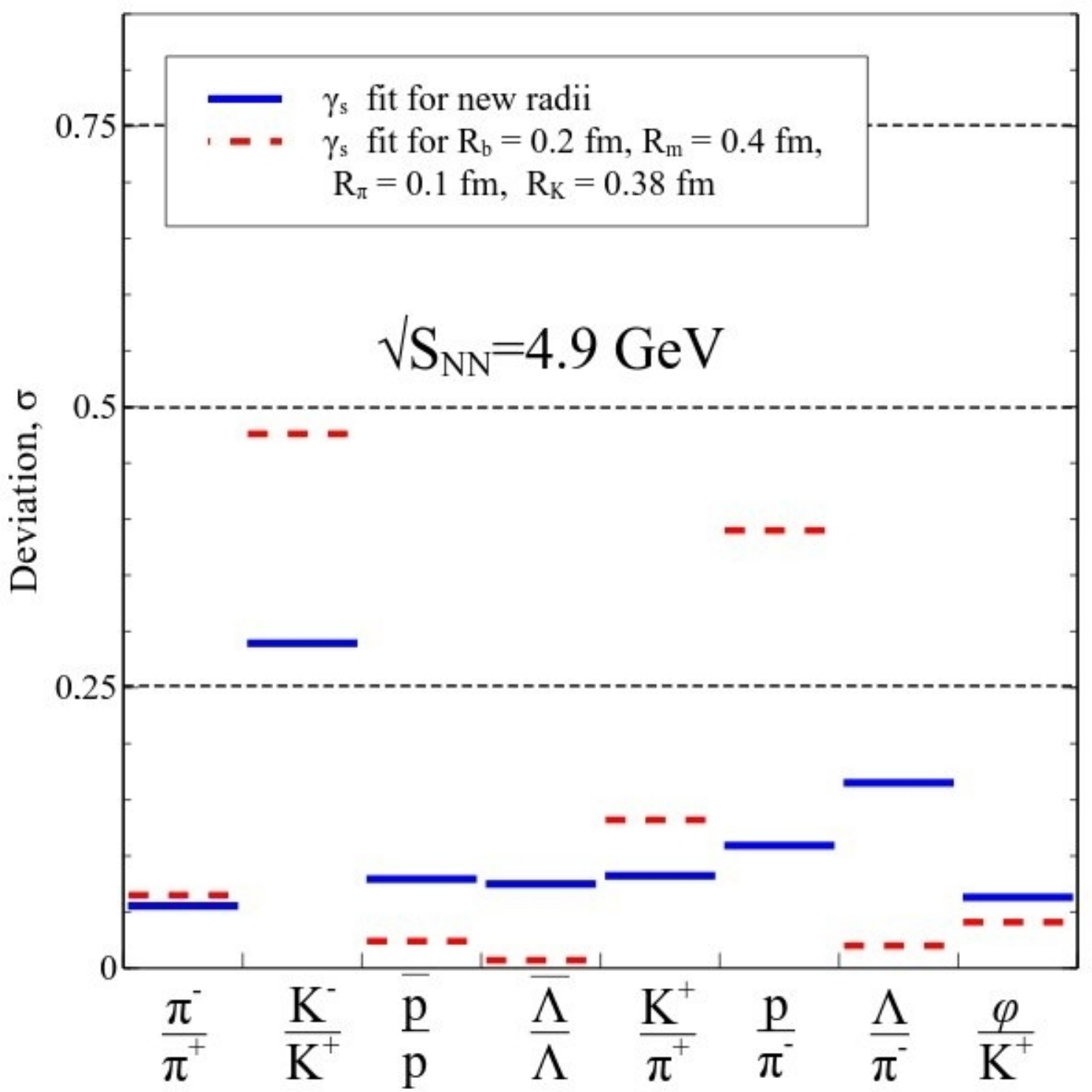}\\[0.5mm]
\includegraphics[width=6.5cm, height=6.15cm]{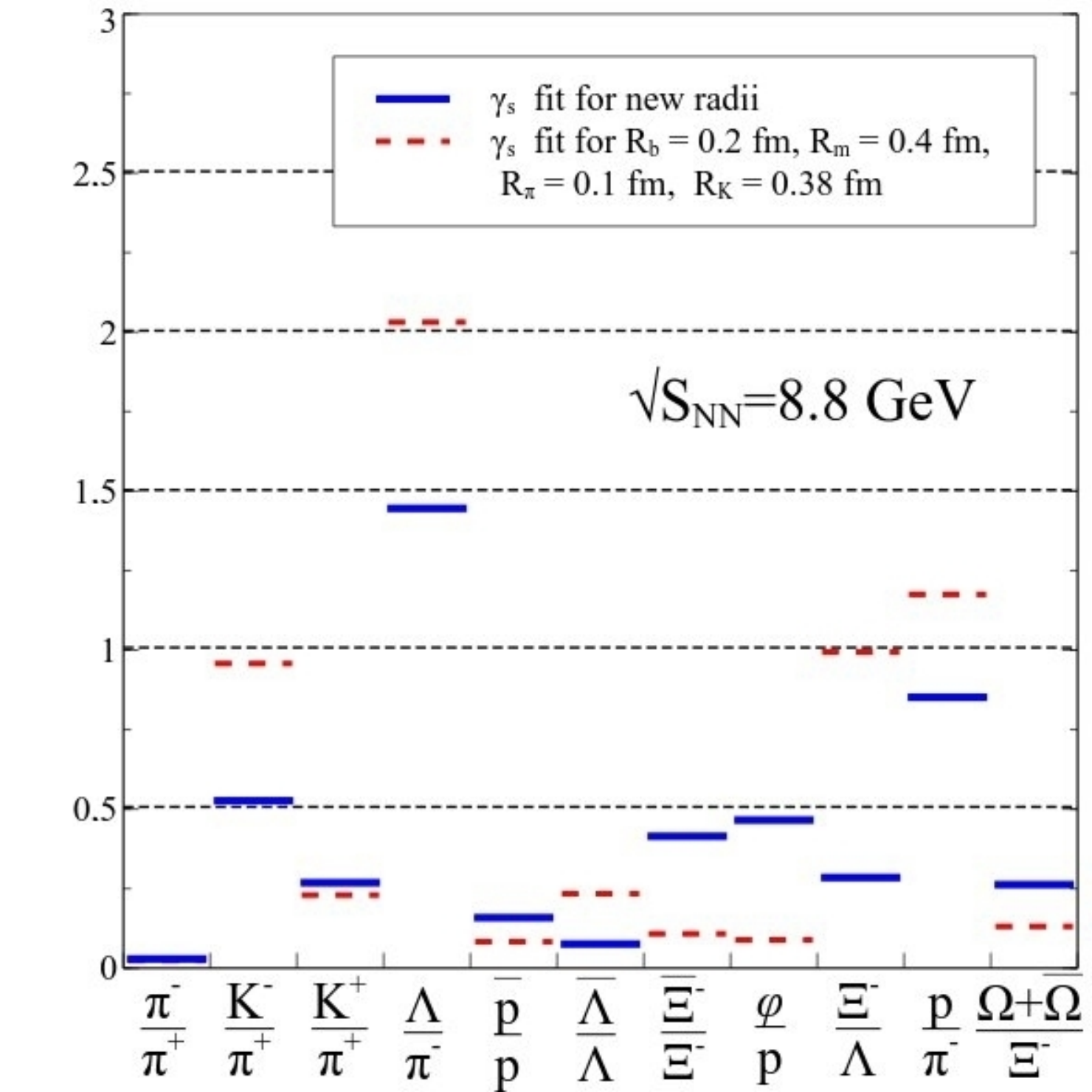}\\[0.5mm]
\includegraphics[width=6.5cm, height=6.15cm]{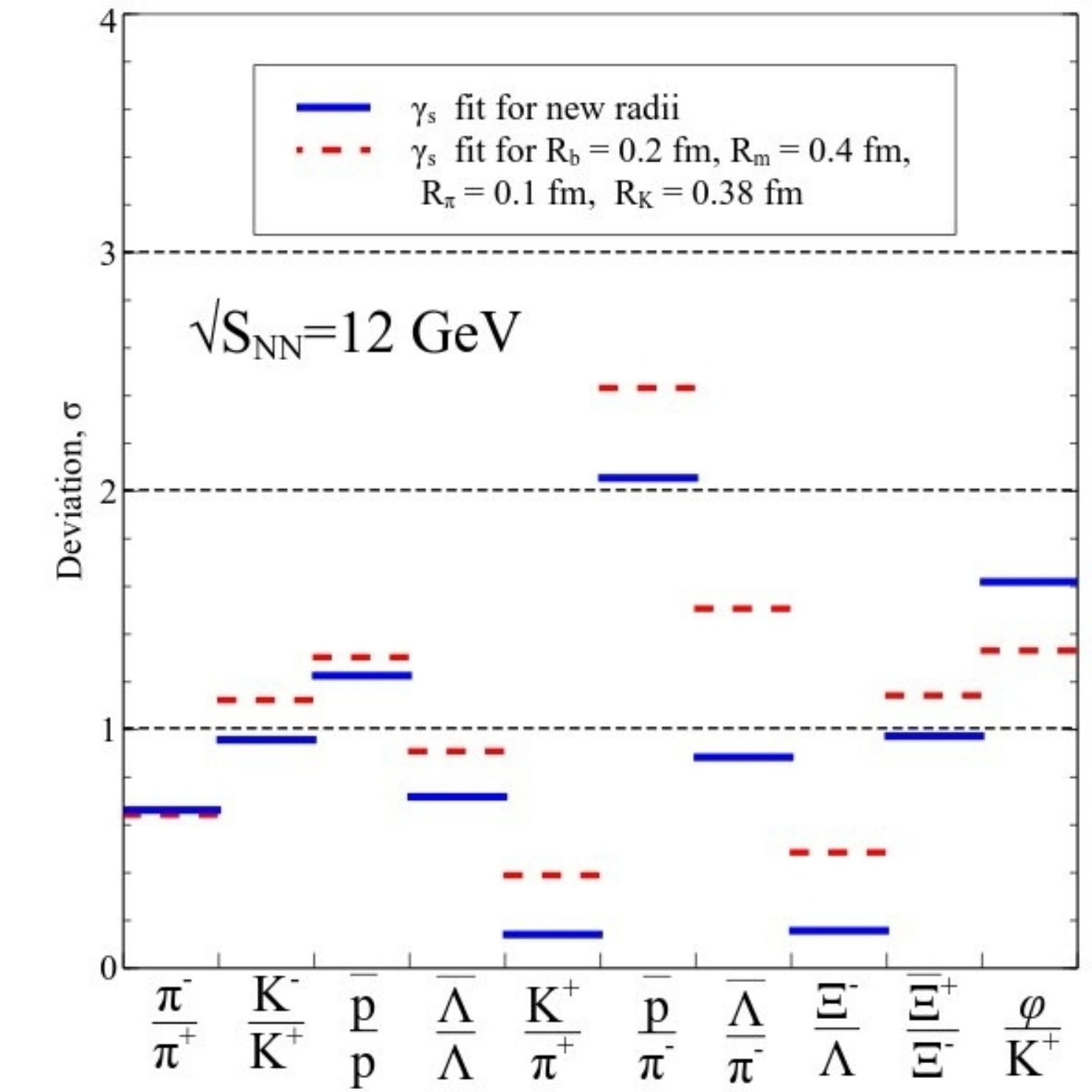}
\vskip-3mm\caption{Relative deviations of theoretical hadron
multiplicity descriptions obtained for the free $\gamma_{s}$ fit
parameter from their experimental values reckoned in the
experimental error $\sigma$ units and their comparison with the
results of fitting in work \cite{KABugaev:Horn2013} also obtained
for the free $\gamma_{s}$ parameter }\label{fig_SagunV}
\end{figure}

Considering the quality of the description of particle ratios for
the fitting with $\gamma_{s}$, we may assert that a general
improvement of the description of particle multiplicities is
observed, especially those including $\Lambda$ hyperons.\,\,For the
energy $\sqrt{s_{NN}}=4.9~\mathrm{GeV}$, unlike the previous fitting
result obtained for $\gamma_{s}=1$, the descriptions of the
$K^{+}/\pi^{+}$, $K^{-}/K^{+}$ and $p/\pi^{-}$ ratios are improved,
which is an additional argument for the necessity of using the
$\gamma_{s}$ factor.\,\,Figure~4 gives some examples of the improved
theoretical description of hadron multiplicities at
$\sqrt{s_{NN}}=4.9$, 8.8 and 12~GeV.

An important result is the improvement of the description of the
$\sqrt {s_{NN}}$ dependences for the ratios that are the most
problematic in the HRGM.\,\,For instance, the description accuracy
for the dependence of the $\Lambda/\pi^{-}$ ratio on the collision
energy becomes higher for two fitting procedures~-- namely, for
$\gamma_{s}=1$ ($\chi^{2}/dof=14.48/12$) and for $\gamma_{s}$ taken
as a free parameter ($\chi^{2}/dof=10.22/12$)~-- in comparison with
the highest accuracy of the previous description attained in work
\cite{BugaevEPL:13} ($\chi^{2}/dof=14.85/12$).\,\,Figure~5
illustrates the result of high-quality fit carried out for the
dependence of the $\bar{\Lambda}/\pi^{-}$ ratio on the collision
ener\-gy.\,\,The corresponding accuracy $\chi^{2}/dof=6.49/8$ was
obtained for the free parameter $\gamma _{s}$, which is better than
the accuracy $\chi^{2}/dof=10/8$ obtained in work
\cite{BugaevSFO13}.\,\,The fit of the Strangeness Horn shows an
insignificant worsening in comparison with its best description
($\chi^{2}/dof=1.5/14$) in work \cite{BugaevSFO13}; however, the
description quality remains high enough, because it can completely
describe the dependence maximum.\,\,The description accuracy for the
$\sqrt{s_{NN}}$ dependence of the $K^{+}/\pi^{+}$ ratio amounts to
$\chi^{2}/dof=$ $=7.25/14$ in the case of the fixed $\gamma_{s}$
factor and to $\chi^{2}/dof=3.92/14$ in the case of the free
$\gamma_{s}$ pa\-rameter.

Despite that the ratio $\chi^{2}/dof$ only slightly differs between
the cases with the fixed or free parameter $\gamma_{s}$~-- namely,
1.09 and 0.95, respectively~-- the $\chi^{2}$-value itself decreased
from 75 to 52, which testifies to a substantial improvement of the
data description.\,\,Hence, a conclusion follows that, in the cases
when the difference between the values of $\chi^{2}/dof$ is
insignificant, it is necessary to introduce an additional criterion
of the data description quality.\,\,As such a criterion, the
description quality of $\Lambda/\pi^{-}$ and $\bar{\Lambda}/\pi^{-}$
multiplicity ratios was selected.\,\,The appreciable improvement of
the description for the most problematic ratios in the HRGM is an
important
argument in favor of $\gamma_{s}$ for the QCD phenomenology.\,\,The $\gamma_{s}%
$-values found at low energies testify to the absence of a chemical equilibrium
between the strange and all other hadrons, which is explained by the
hypothesis of separate chemical freeze-outs \cite{BugaevEPL:13, UJPGamma,
UJPSFO}.

\section{Conclusions}

\label{Conclusions}

The influence of the hadron hard-core radius on the description
quality of experimental hadron yields has been systematically
analyzed in the framework of the multicomponent hadron resonance gas
model.\,\,The hard-core radii found
for baryons, $R_{b}=0.355$~fm, mesons, $R_{m}=0.4$~fm, pions, $R_{\pi}%
=0.1$~fm, kaons, $R_{K}=0.38$~fm and $\Lambda$ hyperons, $R_{\Lambda}%
=0.11$~fm, satisfy the condition of $\chi^{2}$ minimum and provide
the best quality of the description of experimental data.\,\,Those
improvements of the model allowed us to obtain a high-quality fit of
experimental data measured in the collision energy interval from AGS
to RHIC ($\sqrt{s_{NN}}=2.7\div$ $ 200~\mathrm{GeV}$).\,\,The
introduction of the additional hard-core radius for the $\Lambda$
hyperon essentially improved the description of the dependences of
$\Lambda/\pi^{-}$ and $\bar{\Lambda}/\pi^{-}$ ratios on the
collision energy with the accuracy $\chi^{2}/dof=10.22/12$ and
$6.49/8$, respectively.\,\,The description of the Strangeness Horn
shows the absolute correspondence between experimentally and
theoretically determined points with $\chi^{2}/dof=3.92/14$.\,\,At
the collision energies $\sqrt{s_{NN}}=$ $=3.3$, 3.8, 4.9, 6.3, 7.6,
and 8.8~GeV, the value $\gamma_{s}>1$ was obtained, which
corresponds to the strangeness enhancement.\,\,As is seen from
Fig.~3, a local minimum of the parameter $\gamma_{s}$ was revealed
in the intervals of collision energies in the center-of-mass system
from 4.3 to 4.9~GeV, which can be an additional argument in favor of
qualitative changes in the system properties at those energies
\cite{Bugaev:1312a}.\,\,The applied approach made it possible to
describe the ratios between all hadron multiplicities with the
accuracy $\chi ^{2}/dof=52/55\simeq0.95$, which is the best one at
present.

\vskip3mm {\it The author thanks D.R.\,Oliinychenko,
O.I.\,Iva\-nyts\-kyi and, especially, K.O.\,Bugaev for their
invaluable help at the preparation of this publication.\,\,The
author also expresses her gratitude to the Nuclear Physics section
of the National Academy of Sciences of Ukraine for the financial
support in the framework of the program \textquotedblleft
Perspective fundamental researches in nuclear and high-energy
physics\textquotedblright.}


\begin{thebibliography}{99}                                                                                               %
\bibitem {Thermalmodelreview}P.~Braun-Munzinger, K.~Redlich and J.~Stachel,
in \textit{Quark-Gluon Plasma 3,} edited by R.C.~Hwa and X.N.~Wang (World
Scientific, Singapore, 2004), p.~491.

\bibitem {KABAndronic:05}A.~Andronic, P.~Braun-Munzinger and J.~Stachel,
Nucl. Phys.~A \textbf{772}, 167 (2006) and references therein.

\bibitem {KABugaev:Horn2013}K.A.~Bugaev, D.R.~Oliinychenko, A.S.~Sorin and
G.M.~Zinovjev, Eur. Phys. J.~A \textbf{49}, 30 (2013) and references therein.

\bibitem {BugaevEPL:13}K.A.~Bugaev, D.R.~Oliinychenko, J.~Cleymans,
A.I.~Ivanytskyi, I.N.~Mishustin, E.G.~Nikonov and V.V.~Sagun,
Euro\-phys. Lett. \textbf{104}, 22002 (2013).

\bibitem {KABugaev:Stachel}J.~Stachel, A.~Andronic, P.~Braun-Munzinger and
K.~Red\-lich, arXiv: 1311.4662[nucl-th].

\bibitem {Rafelski:gamma}J.~Rafelski, Phys. Lett.~B \textbf{62}, 333 (1991).

\bibitem {Becattini:gammaHIC}F.~Becattini, J.~Manninen and M.~Gazdzicki,
Phys. Rev.~C \textbf{73}, 044905 (2006).

\bibitem {PBM:gamma}P.~Braun-Munzinger, D.~Magestro,K.~Redlich and
J.~Sta\-chel, Phys. Lett.~B \textbf{518}, 41 (2001).

\bibitem {UJPGamma}V.V.~Sagun, D.R.~Oliinychenko, K.A.~Bugaev, J.~Cleymans,
A.I.~Ivanytskyi, I.N.~Mishustin and E.G.~Nikonov, arXiv: 1403.6311[hep-ph].

\bibitem {BugaevIndia}S.~Chatterjee, R.M.~Godbole and S.~Gupta, Phys. Lett.~B
\textbf{727}, 554 (2013).

\bibitem {UJPSFO}D.R.~Oliinychenko, V.V.~Sagun, A.I.~Ivanytskyi and
K.A.~Bugaev, arXiv: 1403.5744[hep-ph].

\bibitem {BugaevtwocompVdW}G. Zeeb, K.A. Bugaev, P.T. Reuter and H.
St\"{o}cker, Ukr. J. Phys. \textbf{53}, 279 (2008).

\bibitem {MultiComp:12}D.R.~Oliinychenko, K.A.~Bugaev and A.S.~Sorin, Ukr. J.
Phys. \textbf{58}, 211 (2013).

\bibitem {BugaevSFO13}K.A.~Bugaev, D.R.~Oliinychenko, V.V.~Sagun,
A.I.~Ivanytskyi, J.~Cleymans, E.G.~Ni\-ko\-nov and G.M.~Zinovjev,
arXiv: 1312.5149 [hep-ph].

\bibitem {KABAndronic:09}A.~Andronic, P.~Braun-Munzinger and J.~Stachel,
Phys. Lett.~B \textbf{673}, 142 (2009).

\bibitem {AGSL3}B.B.~Back \textit{et al.,} Phys. Rev. Lett. \textbf{87},
242301 (2001).

\bibitem {KABugaev:Becattini}F.~Becattini \textit{et al.,} Phys. Rev.~C
\textbf{85}, 044921 (2012).

\bibitem {KABugaev:Becattini13}F.~Becattini \textit{et al.,} Phys. Rev. Lett.
\textbf{111}, 082302 (2013).

\bibitem {SBM}R.~Hagedorn, Nuovo Cim. Suppl. \textbf{3}, 147 (1965).

\bibitem {LeeYang52}T.D.~Lee and C.N.~Yang, Phys. Rev. \textbf{87}, 3 (1952).

\bibitem {Bugaev:1312a}K.A. Bugaev, A.I. Ivanytskyi, D.R.~Oliinychenko,
E.G.~Nikonov, V.V.~Sagun and G.M.~Zinovjev, arXiv: 1312.4367[hep-ph].

\bibitem {THERMUS}S.~Wheaton, J.~Cleymans and M.~Hauer, Comput. Phys. Commun.
\textbf{180}, 84 (2009).

\bibitem {AGSpi1}J.L.~Klay \textit{et al.,} Phys. Rev.~C \textbf{68}, 054905 (2003).

\bibitem {AGSpi2}L.~Ahle \textit{et al.,} Phys. Lett.~B \textbf{476}, 1 (2000).

\bibitem {AGSp1}B.B.~Back \textit{et al.,} Phys. Rev. Lett. \textbf{86}, 1970 (2001).

\bibitem {AGSp2}J.L.~Klay \textit{et al.,} Phys. Rev. Lett. \textbf{88},
102301 (2002).

\bibitem {AGSL}C.~Pinkenburg \textit{et al.,} Nucl. Phys.~A \textbf{698}, 495c (2002).

\bibitem {AGSKas}P.~Chung \textit{et al.,} Phys. Rev. Lett. \textbf{91},
202301 (2003).

\bibitem {AGSphi}B.B.~Back \textit{et al.,} Phys. Rev.~C \textbf{69}, 054901 (2004).

\bibitem {AGSL2}S. Albergo \textit{et al.,} Phys. Rev. Lett. \textbf{88},
062301 (2002).

\bibitem {KABNA49:17a}S.V.~Afanasiev \textit{et al.,} Phys. Rev.~C
\textbf{66}, 054902 (2002).

\bibitem {KABNA49:17b}S.V.~Afanasiev \textit{et al.,} Phys. Rev.~C
\textbf{69}, 024902 (2004).

\bibitem {KABNA49:17Ha}T.~Anticic \textit{et al.,} Phys. Rev. Lett.
\textbf{93}, 022302 (2004).

\bibitem {KABNA49:17Hb}S.V.~Afanasiev \textit{et al.,} Phys. Lett.~B
\textbf{538}, 275 (2002).

\bibitem {KABNA49:17Hc}C.~Alt \textit{et al.,} Phys. Rev. Lett. \textbf{94},
192301 (2005).

\bibitem {KABNA49:17phi}S.V.~Afanasiev \textit{et al.,} Phys. Lett. B
\textbf{491}, 59 (2000).

\bibitem {KABstar:9.2}B.~Abelev \textit{et al.,} Phys. Rev.~C \textbf{81},
024911 (2010).

\bibitem {KABstar:62a}B.~Abelev \textit{et al.,} Phys. Rev.~C \textbf{79},
034909 (2009).

\bibitem {KABstar:130a}J.~Adams \textit{et al.,} Phys. Rev. Lett. \textbf{92},
182301 (2004).

\bibitem {KABstar:130b}J.~Adams \textit{et al.,} Phys. Lett.~B \textbf{567},
167 (2003).

\bibitem {KABstar:130c}C.~Adler \textit{et al.,} Phys. Rev.~C \textbf{65},
041901(R) (2002).

\bibitem {KABstar:200a}J.~Adams \textit{et al.,} Phys. Rev. Lett. \textbf{92},
112301 (2004).

\bibitem {KABstar:200b}J.~Adams \textit{et al.,} Phys. Lett.~B \textbf{612}, 181 (2005).

\bibitem {KABstar:200c}A.~Billmeier \textit{et al.,} J. Phys.~G \textbf{30},
S363 (2004).\vspace*{2mm}
\begin{flushright}
{\footnotesize Received 2014.\\ Translated from Ukrainian by
O.I.~Voitenko}
\end{flushright}
\end{thebibliography}
\end{document}